\documentstyle[prl,aps,twocolumn,floats,epsf,psfig]{revtex}
\begin{document}
\draft
\twocolumn[\hsize\textwidth\columnwidth\hsize\csname @twocolumnfalse\endcsname
\title{Frictional Drag between Two Dilute Two-Dimensional Hole Layers }
\author{R. Pillarisetty, Hwayong Noh, D. C. Tsui, E. P. De Poortere, E. Tutuc,
and M. Shayegan}
\address{Department of Electrical Engineering, Princeton University,
Princeton, NJ 08544}
\date{\today}
\maketitle
\begin{abstract}
We report drag measurements on dilute double layer two-dimensional
hole systems in the regime of $r_{s}=19\sim39$. We observed a
strong enhancement of the drag over the simple Boltzmann
calculations of Coulomb interaction, and deviations from the
$T^{2}$ dependence which cannot be explained by phonon-mediated,
plasmon-enhanced, or disorder-related processes. We suggest that
this deviation results from interaction effects in the dilute
regime.
\end{abstract}
\pacs{73.40.-c,73.21.Ac,73.40.Kp}
]

In recent years, transport properties in dilute two-dimensional
(2D) electron or hole systems have generated much interest from
the possibility that strong interactions between charge carriers
in such systems could stabilize a new quantum state, which can
exhibit unusual transport characteristics. In contradiction to the
scaling theory of localization\cite{aalt}, which predicts only an
insulating state in 2D in the absence of interactions, a metallic
behavior characterized by a decreasing resistivity ($\rho$) with
decreasing temperature ($T$) and an apparent metal-insulator
transition as the carrier density is varied has been observed in
many dilute 2D systems\cite{mit}. Despite numerous experimental
and theoretical studies, there is no conclusive understanding of
the nature of the metallic behavior at this point. It is an
interesting question whether this metallic behavior is related to
any non-Fermi liquid behavior, since the interactions between the
carriers are strong. Studies\cite{proskuryakov} indicate that the
2D system behaves as a Fermi liquid even in this dilute
metal-insulator transition regime by examining the weak
localization and interaction corrections to the conductivity. In
this paper, we present an alternative way of studying the validity
of Fermi liquid theory in this dilute regime: frictional drag
experiment.

Drag measurements, which are performed on double layer structures,
have been utilized to study the interlayer carrier-carrier
interactions\cite{TJ,drag}. These measurements are done by passing
a drive current ($I_{drive}$) through one of the layers, while
measuring the potential difference ($V_{D}$), which arises in the
other layer due to carrier momentum transfer. The drag
resistivity, given by $\rho_{D}=(W/L)V_{D}/I_{drive}$, where $W$
and $L$ are the width and the length of the sample, provides a
direct measure of the interlayer scattering rate in these
structures. The first component of $\rho_{D}$ has been identified
as from the Coulomb interaction, which is characterized by a
quadratic increase of $\rho_{D}$ with $T$ in a simple single
particle picture\cite{TJ,coloumb,halldrag}. Any deviations from
this $T^{2}$ dependence would mean either a non-Fermi liquid
behavior or an additional interaction mechanism. Most previous
drag measurements have been performed on high density samples, and
confirmed the $T^{2}$ dependence when the layer separation was 300
\AA\ and less\cite{halldrag}, while they showed deviations from
this dependence\cite{TJ,phonon,phonon2,german,plasmon} when either
the layer separation is large or the temperature is high. These
deviations, shown as a peak when $\rho_{D}/T^{2}$ is plotted as a
function of $T$, were attributed to either a phonon-mediated
process\cite{TJ,phonon,phonon2,german,phonont} or a plasmon
excitation contribution\cite{plasmon,plasmont}.

\begin{figure}[!t]
\begin{center}
\leavevmode
\hbox{%
\epsfxsize=3.25in
\epsffile{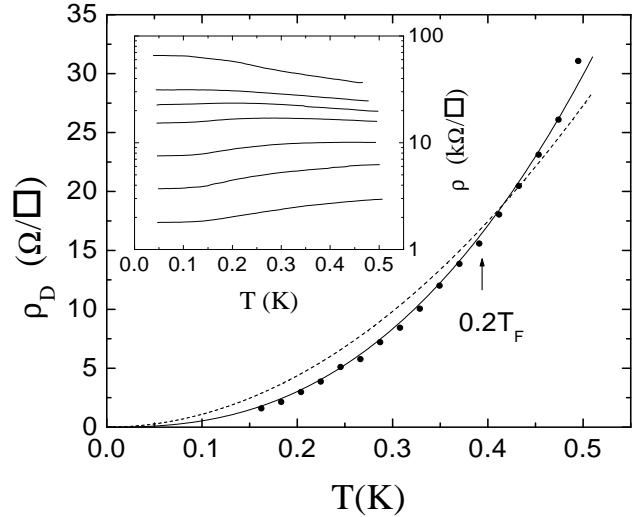}}
\end{center}
\caption{$\rho_{D}$ vs $T$ for sample A. Layer densities are
matched at $p=2.5\times10^{10}$ cm$^{-2}$. Provided are the best
$T^{2}$ (dashed line) and $T^{2.5}$ (solid line) fits. Inset:
$\rho$ vs $T$ for the bottom layer. From top to bottom, $p=$
0.80, 0.85, 0.90, 1.0, 1.2, 1.4, and $2.0\times10^{10}$
cm$^{-2}$.}
\label{1}
\end{figure}

What is of major interest in our study is the drag in the low
density regime with small layer separation in order to investigate
the validity of $T^{2}$ dependence in the $T\rightarrow0$ limit, a
test for the Fermi liquid behavior. In this paper, we present drag
measurements on dilute two dimensional hole systems at layer
densities $(p)$ ranging from $2.5\times10^{10}$ cm$^{-2}$ to
$6\times10^{9}$ cm$^{-2}$.  These are the first reported results,
to our best knowledge, of drag measurements in this low density
range. In addition, the larger effective mass ($m^{*}$) in the
hole samples allows us to explore a much higher $r_{s}$ regime
than has been previously measured in electron samples; $r_{s}$ is
the ratio of interaction energy to Fermi energy and is given by
$r_{s}=(p\pi)^{-1/2}m^{*}e^{2}/4\pi\hbar^{2}\epsilon\epsilon_{0}$,
where $\epsilon$ is the dielectric constant. The $r_{s}$ values of
the 2D systems which exhibit metal-insulator transition is usually
around 10 or higher at the transition. Using the hole effective
mass for our system $m^{*}=0.38m_{e}$, we find that the $r_{s}$
values for our measurements range from 19 to 39. This is more than
an order of magnitude larger than the $r_{s}$ values for the
lowest density electron systems which have been
measured\cite{halldrag}. From the $T$ dependence of $\rho$ taken
from one of the layers, we observed a clear metal-insulator
transition at $p\sim8.5\times10^{9}$ cm$^{-2}$ and
$\rho\sim30k\Omega/\Box$ (inset of Fig.~\ref{1}).

Our drag measurements in this regime show that, for
$T\le0.5T_{F}$, the drag exhibits a slightly stronger than $T^{2}$
dependence, where $T_{F}$ is the Fermi temperature. The magnitude
of the drag is much larger than the simple Boltzmann calculations
predict\cite{coloumb}. As
the temperature is further increased we find a crossover to a
sub-linear dependence, and when $\rho_{D}/T^{2}$ is plotted as a
function of $T$, the data exhibits a peak similar to that in a
phonon-mediated\cite{TJ,phonon,phonon2,german,phonont} or
plasmon-enhancement process\cite{plasmon,plasmont}. However, from
the dependence on the density ratio of the two layers, we observe
no evidence of such processes in our systems. In addition, we find
our data to be inconsistent with previous disorder enhanced
studies of the drag\cite{disorder}. These findings lead us to
believe that interaction effects are behind the observed
enhancement.

The samples used in this experiment are Si $\delta$-doped
AlGaAs/GaAs/AlAs double quantum well structures, which were grown
by molecular beam epitaxy on the (311)A surface. Most of this
paper presents data taken from a low density double layer hole
sample, which we will refer to as sample A. We also provide data
from an intermediate density double layer hole sample, which we
refer to as sample B. The structure of sample A (sample B)
consists of two 150 (175) \AA\ GaAs quantum wells separated by a
150 (100) \AA\ AlAs barrier, corresponding to a center to center
layer separation ($d$) of 300 (275) \AA. The grown densities in each
layer are $2.5\times10^{10}$ cm$^{-2}$ for sample A, and
$7.0\times10^{10}$ cm$^{-2}$ for sample B. For sample A, the
mobility of each layer is roughly $1.5\times10^{5}$ cm$^{2}$/Vs
for the grown densities at 300 mK. For sample B, the mobilities
are, for the grown densities, $5.7\times10^{5}$ cm$^{2}$/Vs and
$7.7\times10^{5}$ cm$^{2}$/Vs for the top and bottom layer
respectively, at 300 mK. The samples were processed using a
selective depletion scheme allowing independent contact to each of
the two layers\cite{IC}. Both layer densities are independently
tunable with the aid of gates evaporated on both sides of the
sample.

The data presented in this paper were taken in a $^{3}$He
refrigerator and in a dilution refrigerator. The layer densities
were determined by independently measuring Shubnikov-de Haas
oscillations of each layer. Drive currents between 500 pA to 10 nA
were passed through one of the layers, while the drag signal was
measured in the other layer, using standard lock-in techniques at
1 to 4 Hz. We performed all the standard consistency checks on our
measurements\cite{TJ}. The grounding to the drag layer was swapped
between voltage contacts and the signals matched. The drag and
drive layers were swapped and showed good agreement. The drag
signal showed a linear variation with the drive current. Also, the
out of phase signal was monitored to ensure that capacitive
coupling between the layers was not affecting our measurement.  It
should be pointed out that three working samples were made from
wafer A and two from wafer B. All of the samples from each wafer
showed reasonable agreement with each other. However, data are
presented from only one sample from each wafer.

In Fig.~\ref{1}, we present $\rho_{D}$ vs $T$ for sample A at
$p=2.5\times10^{10}$ cm$^{-2}$. The temperatures in this plot
range from $0.08T_{F}$ to $0.25T_{F}$. When a trial fit is made
with $T^{2}$(dashed line), the data does not show good agreement.
Instead, when we fit the data with $T^{2.5}$(solid line), a better
result is obtained. When the exponent is varied as a free
parameter, we also find a value close to 2.5. It is clear that at
low temperatures, $\rho_{D}$ shows a stronger than $T^{2}$
dependence. In addition, we find a very large enhancement of
$\rho_{D}$. For this density, our measurements are greater than
200 times that calculated from the Boltzmann
equation\cite{coloumb} (valid for $k_{F}d\gg 1$ and $T\ll T_{F}$,
with $k_{F}d=1.2$ and $T<0.25T_{F}$ in our measurement)
and roughly 5 orders of magnitude larger
than the original drag experiment involving high density electrons
($n=1.5\times10^{11}$ cm$^{-2}$) with $d=375$ \AA\ 
($\rho_{D}=0.5 m\Omega/\Box$ at 500 mK)\cite{TJ}. More recent measurements
on bilayer electron systems\cite{halldrag}, which have similar
$d$ to our samples, showed a nearly
quadratic temperature dependence for densities ranging from
$5.3\times10^{10}$ cm$^{-2}$ to $2.5\times10^{10}$ cm$^{-2}$,
which are slightly larger than ours samples. For the largest
density, they measured $\rho_{D}/T^{2}\sim0.4\Omega/\Box K^{2}$,
which is only four times larger than that expected from a
Boltzmann calculation ($k_{F}d=1.6$, which is comparable to our sample).
It is evident that
Boltzmann transport theory fails to provide either the proper
temperature dependence or the correct magnitude for measurement in
our samples.

\begin{figure}[!b]
\begin{center}
\leavevmode
\hbox{%
\epsfxsize=3.25in
\epsffile{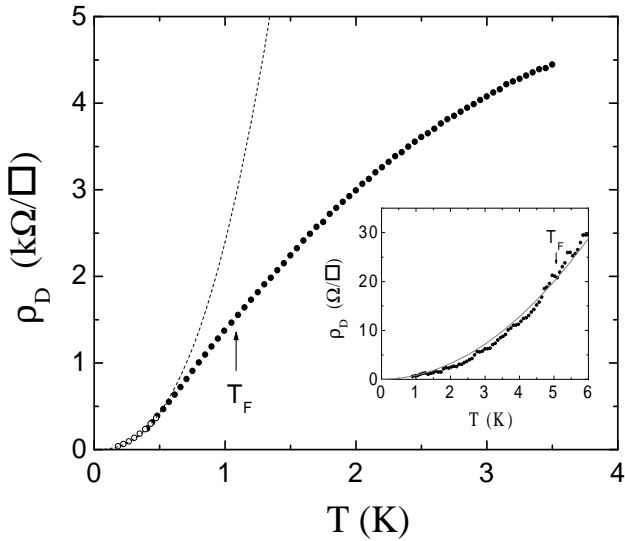}}
\end{center}
\caption{$\rho_{D}$ vs $T$ for sample A. Layer densities are
matched at $p=1.5\times10^{10}$ cm$^{-2}$. Data measured in the
dilution refrigerator are shown with open circles. Also shown is
the best $T^{2.5}$ fit of these points. Inset: $\rho_{D}$ vs $T$
for sample B, whose layer densities are matched at
$p=7.0\times10^{10}$ cm$^{-2}$. Also shown is the best $T^{2}$ fit
of the data. The Fermi temperature is indicated on both plots.}
\label{2}
\end{figure}

In Fig.~\ref{2}, we present the complete temperature dependence of
$\rho_{D}$ for matched layer densities at $p=1.5\times10^{10}$
cm$^{-2}$ taken from sample A. Although the data points on the low
temperature side are again well described by a $T^{2.5}$ fit
(dashed line), as the temperature is further increased we find a
crossover to a sub-linear dependence at approximately $0.5T_{F}$.
Also, we find the magnitude at this density to be roughly 500
times larger than the Boltzmann calculation\cite{coloumb}. For
comparison we also present intermediate density data taken from
sample B. For this sample $d=275$ \AA,
so comparison is reasonable. For matched layer densities,
$p=7\times10^{10}$ cm$^{-2}$, the data exhibits a nearly quadratic
temperature dependence, and has a magnitude about 20 times larger
than that calculated from the Boltzmann equation\cite{coloumb}. At
this intermediate density, we find no evidence that the system
cannot be described by Fermi liquid theory. Note that this data
shows no significant deviation from $T^2$ up to and above the
Fermi temperature of about 5 K. This is in contrast to the low
density sample (sample A), which clearly exhibits non-quadratic
behavior at temperatures far below $T_{F}$.

\begin{figure}[!b]
\begin{center}
\leavevmode
\hbox{%
\epsfxsize=3.25in
\epsffile{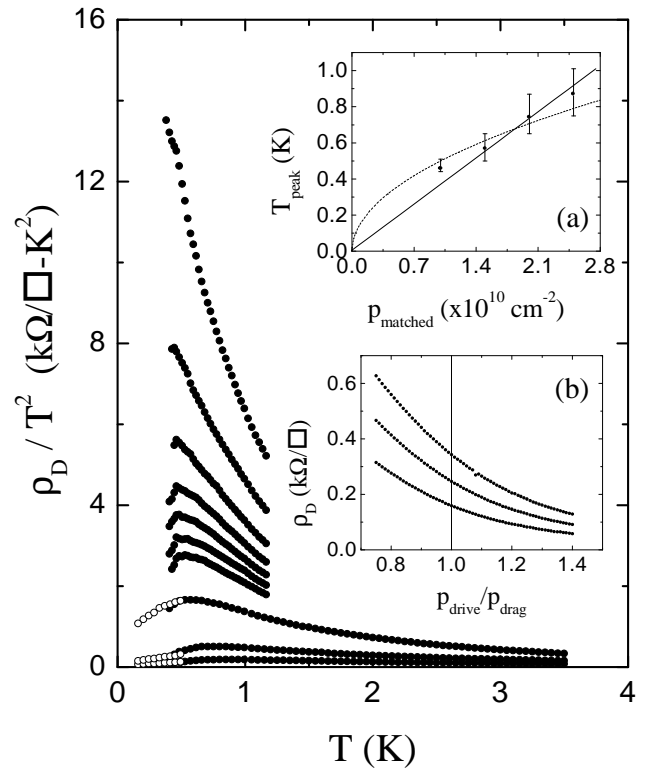}}
\end{center}
\caption{$\rho_{D}/T^{2}$ vs $T$ for sample A. Data measured in
the dilution refrigerator are shown as open circles. From top to
bottom, the layer densities($p_{drive}/p_{drag}$) in $10^{10}$
cm$^{-2}$ are 0.6/1.0, 0.8/1.0, 1.0/1.0, 1.0/1.3, 1.1/1.4,
1.2/1.4, 1.5/1.5, 2.0/2.0, 2.5/2.5. Inset: (a) Peak position
temperature vs matched layer density. Also provided are the best
linear (solid line) and square root (dashed line) fits of the
data. (b) $\rho_{D}$ vs density ratio at (from top to bottom) $T=$
860, 730, and 600 mK. The drag layer is fixed at
$2.0\times10^{10}$ cm$^{-2}$, while the drive layer is swept from
2.4 to 1.5$\times10^{10}$ cm$^{-2}$, passing through the matched
density point indicated by the solid line.}
\label{3}
\end{figure}

The nature of this deviation from the $T^2$ dependence in the
dilute regime is of much importance, since it possibly could be
representative of non-Fermi liquid behavior of the strongly
interacting charge carriers. On the other hand, it is also
possible that the system can still be described as a Fermi liquid
and the deviations from $T^2$ arise from the activation of some
other interlayer momentum transfer mechanism. We approach this
question by examining the deviation from the simple quadratic
dependence  more closely. In Fig.~\ref{3}, we present the $T$
dependence of $\rho_{D}/T^{2}$. Data is shown for layer densities
from $2.5\times10^{10} $cm$^{-2}$ to below $1\times10^{10}$
cm$^{-2}$. The deviation is clearly demonstrated by a local maxima
around a characteristic crossover temperature, similar to previous
drag measurements involving phonon
mediated\cite{TJ,phonon,phonon2,german} and plasmon enhanced
processes\cite{plasmon}.

For electron scattering mediated by phonons, this deviation arises
from scattering processes involving 2$k_{F}$ phonons, and the
expected peak in $\rho_{D}/T^{2}$ vs $T$ was observed in higher
density samples near the Bloch-Gr\"{u}neisen temperature
($T_{BG}$), given by $k_{B}T_{BG}=2\hbar k_{F}c_{s}$, where
$c_{s}$ is the sound velocity in the barrier\cite{phonon,phonon2,german}. 
For our sample,
$T_{BG}$ ranges from 1 to 2 K, and the phonon-mediated process is
certainly a possibility. In previous experiments on electron
samples, a second peak in $\rho_{D}/T^{2}$ was observed at still
higher temperatures near $0.5T_{F}$, which arises from a plasmon
excitation\cite{plasmon}. Since $0.5T_{F}$ in our low density
sample ranges from 500 mK to 1.5 K, we expect the peak positions
arising from both phonon and plasmon processes to fall in roughly
the same range. Therefore, we look at the density dependence of
the peak position to determine which of these could be
contributing to $\rho_{D}$.

In the inset (a) of Fig.~\ref{3}, we plot the peak position versus
the matched layer density. It is clear that the peak position
shifts to lower temperature as the density is lowered. We expect,
for phonon mediated processes, this peak should scale with the
square root of the matched density. On the other hand for plasmon
enhanced processes, the peak should scale with the matched layer
density. To determine whether the peak in the scaled drag
resistivity could be arising from phonon or plasmon processes, we
provide the best square root fit (phonon) and the best linear fit
(plasmon) of the data. Considering the error bar associated with
the determination of peak position and the narrow range of data,
we found that the precise density dependence of this peak position
cannot be specified accurately.

Yet, regardless of whether phonon or plasmon processes contribute
to the drag resistivity, this enhancement should be strongest when
the layer densities are matched. In the former, phonon exchange
between carriers in each layer mediates the scattering process,
and one would expect the enhancement of the drag resistivity to be
largest when the 2$k_{F}$ phonons in each layer are identical.
Drag measurements on the high density electron samples exhibited a
sharp local maxima when the densities were
matched\cite{phonon,phonon2}. Similar to the phonon case, the
plasmon enhancement of the drag resistivity was also shown to be
maximized when the layer densities were matched\cite{plasmon}. In
the inset (b) of Fig.~\ref{3}, we present measurements of
$\rho_{D}$ versus density ratio of the two layers. The drive layer
density was slowly changed for a fixed drag layer density of
$2.0\times10^{10}$ cm$^{-2}$, and the resulting drag signal was
measured. The measurements were performed near the peak
temperature for matched density of $2.0\times10^{10}$ cm$^{-2}$.
We notice that the density dependence is monotonic and we see no
structure at all anywhere around the matched density point.
Although we may not expect a strong peak at the matched density
because the Fermi temperature difference between the two layers
are about the same as the measurement temperature in this plot,
we note that the previous plasmon-enhanced drag measurements\cite{plasmon} 
showed a peak under similar conditions. If the thermal broadening
makes the peak invisible for the phonon-mediated process, we expect
that the peak in the temperature dependence in Fig.~\ref{3} would 
be broadened significantly as well, which is not the case in our data.

From the density dependence data, we find our measurements are
inconsistent with the results of previous drag experiments
involving phonons and plasmons. In fact, no evidence of either of
these were seen in our intermediate density sample (Fig.~\ref{2}
inset), and there is no reason to believe that these mechanisms
should become much stronger as the density is lowered. On the
other hand, we know that as the density is tuned into the dilute
regime both disorder and interaction effects play an important
role in charge transport. In the diffusive limit as
$T\rightarrow0$, the disorder is expected to enhance the drag, and
change the temperature dependence from the expected $T^2$ to
$-T^2ln{T}$\cite{disorder}. Investigating the possibility that the
deviation from the $T^2$ dependence in our low temperature data
could be due to diffusive behavior, we found that the data could
not be fit by $-T^2ln{T}$. In addition, the inset (a) of
Fig.~\ref{3} shows that the boundary of the regime of the
enhancement to the $T^2$ dependence shifts to lower temperature as
the density and, in turn, the mobility is lowered. This is in
contrast to the expectation that the boundary of the diffusive
behavior should scale with the disorder temperature of the system,
given by $\hbar/\tau$, where $\tau$ is the transport scattering
time, and it should shift to higher temperatures as the mobility
is lowered. These inconsistencies lead us to believe that no
signatures of the disorder are evident in our data.

In conclusion, we find ourselves unable to explain the drag
measurements in this dilute regime by accounting for additional
corrections due to phonon, plasmon or disorder enhanced processes.
We are therefore led to believe that the characteristics observed
in the data are due to interaction effects.
Calculations\cite{correlations} have shown that for electron
systems with intermediate densities, $\rho_{D}$ still exhibits a
$T^2$ dependence at low temperatures and is enhanced by an order
of magnitude over the Boltzmann calculations by accounting for
intralayer correlations. However, we find that this calculation
can only partly explain our data since we observe a much larger
enhancement and a greater than $T^2$ dependence. We therefore find
it still unclear whether the strong interaction between carriers
simply adds more corrections which enhance and alter the
temperature dependence of the drag, or if the interaction
stabilizes a new state which exhibits non-Fermi liquid behavior.

We acknowledge S. Das Sarma, S. Simon, and T. J. Gramila for
useful discussions related to the data. This work is supported by
the NSF, and a DURINT grant through the ONR.


\vspace{-0.18in}
\begin{references}
\vspace{-0.68in}

\bibitem{aalt} E. Abrahams, P.W. Anderson, D.C. Licciardello, and T.V.
Ramakrishnan, Phys. Rev. Lett. {\bf 42}, 673 (1979).
\bibitem{mit} For a recent review see B.L. Altshuler, D.L. Maslov, and V.M.
Pudalov, Physica E (Amsterdam) {\bf 9}(2), 209 (2001); E.
Abrahams, S.V. Kravchenko, and M.P. Sarachik, Rev. Mod. Phys. {\bf
73}, 251 (2001).
\bibitem{proskuryakov} M.Y. Simmons {\it et al.}, Phys. Rev. Lett.
{\bf 84}, 2489 (2000).
\bibitem{TJ} T.J. Gramila {\it et al.}, Phys. Rev. Lett. {\bf 66},
1216 (1991).
\bibitem{drag} See for a review A.G. Rojo, J. Phys.: Condens. Matter {\bf 11},
R31 (1999).
\bibitem{coloumb} A. Jauho and H. Smith, Phys. Rev. B {\bf 47}, 4420 (1993).
\bibitem{halldrag} M. Kellogg {\it et al.}, cond-mat/0108403.
\bibitem{phonon} T.J. Gramila {\it et al.}, Phys. Rev. B {\bf 47}, 12957
(1993).
\bibitem{phonon2} H. Rubel {\it et al.}, Semicond. Sci. Technol. {\bf 10},
1229 (1995); H. Noh {\it et al.}, Phys. Rev. B {\bf 59}, 13114
(1999).
\bibitem{german} C. J\"{o}rger {\it et al.}, Physica E {\bf 6}, 598 (2000).
\bibitem{phonont} M. B\o nsager {\it et al.}, Phys. Rev. B {\bf 57},
7085 (1998).
\bibitem{plasmon} N.P.R. Hill {\it et al.}, Phys. Rev. Lett.
{\bf 78}, 2204 (1997); H. Noh {\it et al.}, Phys. Rev. B {\bf 58},
12621 (1998).
\bibitem {plasmont} K. Flensberg and B.Y.-K. Hu, Phys. Rev. Lett. {\bf 73},
3572 (1994).
\bibitem{disorder} L. Zheng and A.H. MacDonald, Phys. Rev. B {\bf 48}, 8203
(1993).
\bibitem{IC} J.P. Eisenstein, L.N. Pfieiffer, and K.W. West, Appl. Phys.
Lett. {\bf 57}, 2324 (1990).
\bibitem{correlations} L. \'{S}wierkowski, J. Szyma\'{n}ski, and Z.W. Gortel,
Phys. Rev. B {\bf 55}, 2280 (1997).

\end{references}
\end{document}